%% file: main.tex
\documentclass[conference]{IEEEtran}

\usepackage{cite}
\usepackage{amsmath,amssymb,amsfonts}
\usepackage{algorithmic}
\usepackage{graphicx}
\usepackage{textcomp}
\usepackage{xcolor}
\usepackage{booktabs}
\usepackage{enumitem}
\usepackage{color}
\usepackage{bm}
\usepackage{subcaption}
\usepackage{url}

\def\BibTeX{{\rm B\kern-.05em{\sc i\kern-.025em b}\kern-.08em
    T\kern-.1667em\lower.7ex\hbox{E}\kern-.125emX}}
\begin{document}

\title{Enhanced Control for Diffusion Bridge in Image Restoration}

\author{\IEEEauthorblockN{1\textsuperscript{st} Conghan Yue \quad 2\textsuperscript{nd} Zhengwei Peng \quad 3\textsuperscript{rd} Junlong Ma \quad 4\textsuperscript{th} Dongyu Zhang}
\IEEEauthorblockA{\textit{School of Computer Science, Sun Yat-Sen University}}
}

\maketitle

\begin{abstract}
Image restoration refers to the process of restoring a damaged low-quality image back to its corresponding high-quality image. Typically, we use convolutional neural networks to directly learn the mapping from low-quality images to high-quality images achieving image restoration. Recently, a special type of diffusion bridge model has achieved more advanced results in image restoration. It can transform the direct mapping from low-quality to high-quality images into a diffusion process, restoring low-quality images through a reverse process. However, the current diffusion bridge restoration models do not emphasize the idea of conditional control, which may affect performance. This paper introduces the ECDB model enhancing the control of the diffusion bridge with low-quality images as conditions. Moreover, in response to the characteristic of diffusion models having low denoising level at larger values of \(\bm t \), we also propose a Conditional Fusion Schedule, which more effectively handles the conditional feature information of various modules. Experimental results prove that the ECDB model has achieved state-of-the-art results in many image restoration tasks, including deraining, inpainting and super-resolution. Code is avaliable at \url{https://github.com/Hammour-steak/ECDB}.
\end{abstract}


\input{content/Intro}
\input{content/RelatedWork}
\input{content/Preliminary}
\input{content/ECDB}

\input{content/Experiment}

\input{content/Conclusion}

\bibliographystyle{IEEEtran}

\bibliography{mybibfile}

\end{document}

%% file: content/Intro.tex
\section{Introduction}

\begin{figure}[!ht]
  \centering
  \includegraphics[width=0.5\textwidth]{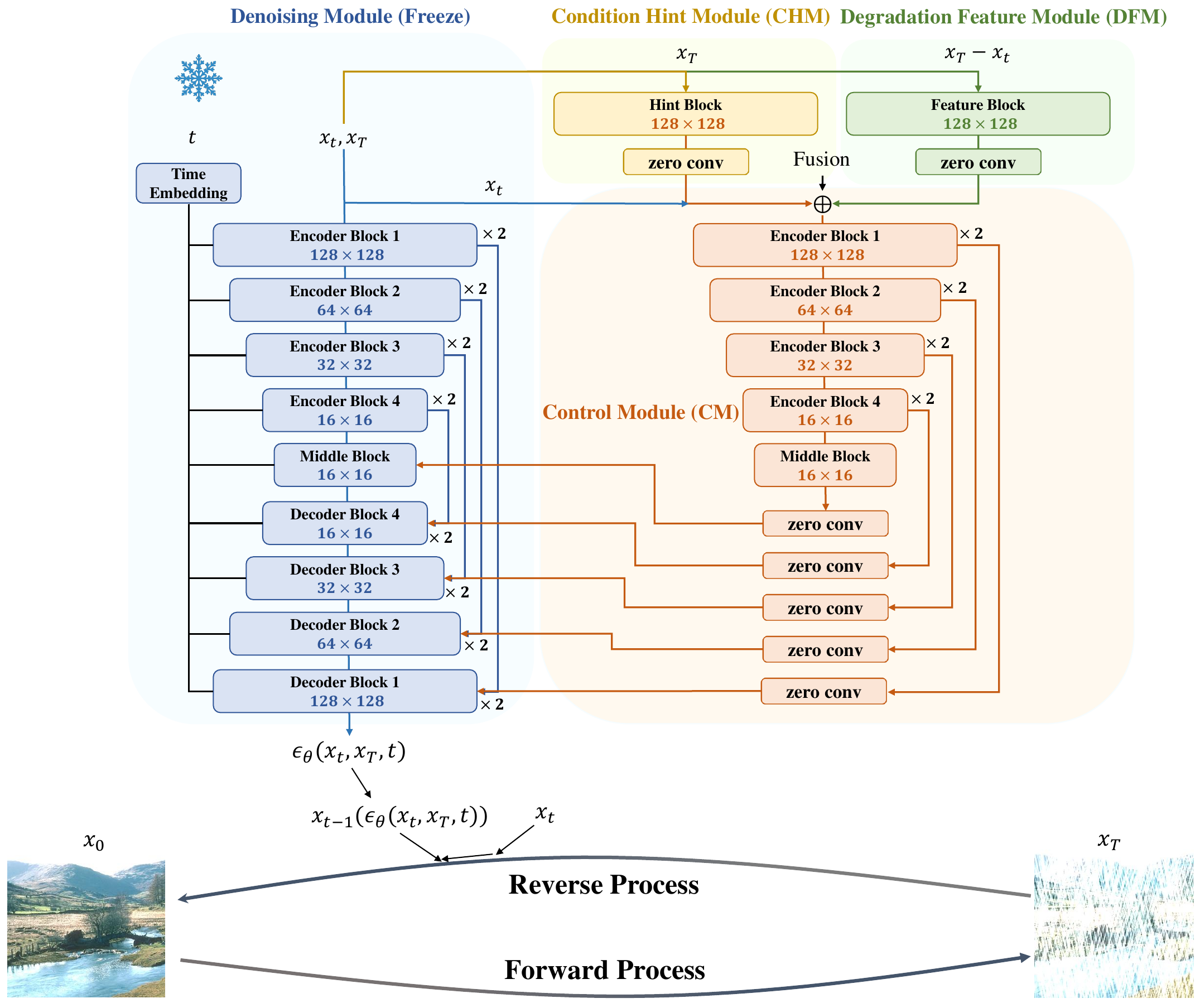}
  \caption{The overview of the proposed ECDB}
  \label{ECDB}
\end{figure}

Image Restoration \cite{banham1997digital, zhou1988image} is pivotal in low-level computer vision, aiming to enhance images impaired by diverse disturbances including noise \cite{tian2023multi}, missing \cite{chen2023ffti, corneanu2024latentpaint}, raindrop \cite{tao2024mffdnet}, resolution decline \cite{chen2023dual, yue2024resshift}, and other distortions. 


In recent years, with diffusion models \cite{sohl2015deep, ho2020denoising, songscore} making remarkable performance in generative tasks, numerous methods \cite{kawar2021snips, chungdiffusion, chung2022improving, li2023bbdm} have leveraged diffusion models to achieve favorable results in image restoration tasks. One of the most noteworthy developments is the diffusion bridges \cite{liu2023i2sb, shi2024diffusion, tongsimulation, zhoudenoising}, which ingeniously integrates the end-to-end training paradigm of CNN models with the denoising concept of diffusion models, establishing a point-to-point diffusion process between high-quality and low-quality images.

A diffusion bridge model that performs well in general image restoration tasks is GOUB \cite{yue2023image}, which involves applying Doob's \textit{h}-transform to the Generalized Ornstein-Uhlenbeck (GOU) process, resulting in a point-to-point diffusion bridge model. Furthermore, GOUB's superiority over other diffusion bridge models has been empirically demonstrated in terms of theoretical advantages. 
However, GOUB solely focuses on controlling the diffusion process by the low-quality image as the condition, neglecting the conditions on the architectural configuration of the model, which in practice constrains its performance capabilities.

In this paper, we introduce ECDB model, which incorporates low-quality images as conditions into the architecture allowing for more comprehensive control over the predicted noise. The ECDB is primarily composed of four modules: the Denoising Module (DM), the Condition Hint Module (CHM), the Degradation Feature Module (DFM) and the Control Module (CM). DM represents the original model of GOUB, the CHM processes the conditions extracting features such as pixels or colors, and the DFM is purpose to extract degradation features. In addition, in light of the distinctive properties of low denoising level at larger values of $t$ in the diffusion process, we introduce a Conditional Fusion Schedule to fuse various conditional features. Ultimately, CM processes the fused features to produce enhanced control information and generates predicted noise with DM, thereby improving the quality of image restoration.
Our main contributions can be summarized as follows:
\begin{itemize}
	\item We introduce the ECDB model, composed of the DM, CHM, DFM, and CM, which strengthens the conditional control over the model and boasts significant portability. Furthermore, due to the necessity of updating only a part of the parameters, it achieves high training efficiency.
	\item Considering the characteristic of low denoising level at larger values of $t$ in the diffusion process, we propose a Conditional Fusion Schedule that more effectively provides conditional control in the denoising module.
	\item The ECDB model has achieved state-of-the-art results in numerous image restoration tasks, such as inpainting, deraining and super-resolution, and has also demonstrated competitive performance on real-world datasets.
\end{itemize}

\begin{figure}[!t]
    \centering
    \begin{subfigure}[b]{0.22\textwidth}
        \centering
        \raisebox{3mm}{\includegraphics[width=\textwidth]{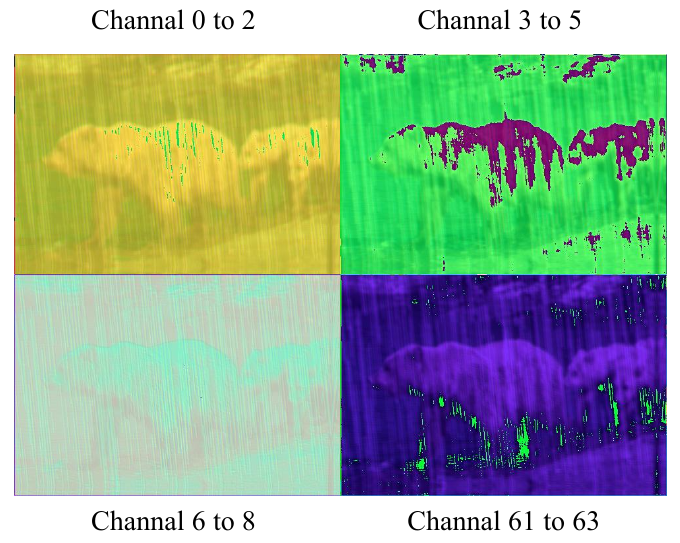}}
        \caption{Visualization of CHM features on partial channels}
        \label{CHM}
    \end{subfigure}
    \hfill
    \begin{subfigure}[b]{0.23\textwidth}
        \centering
        \includegraphics[width=\textwidth]{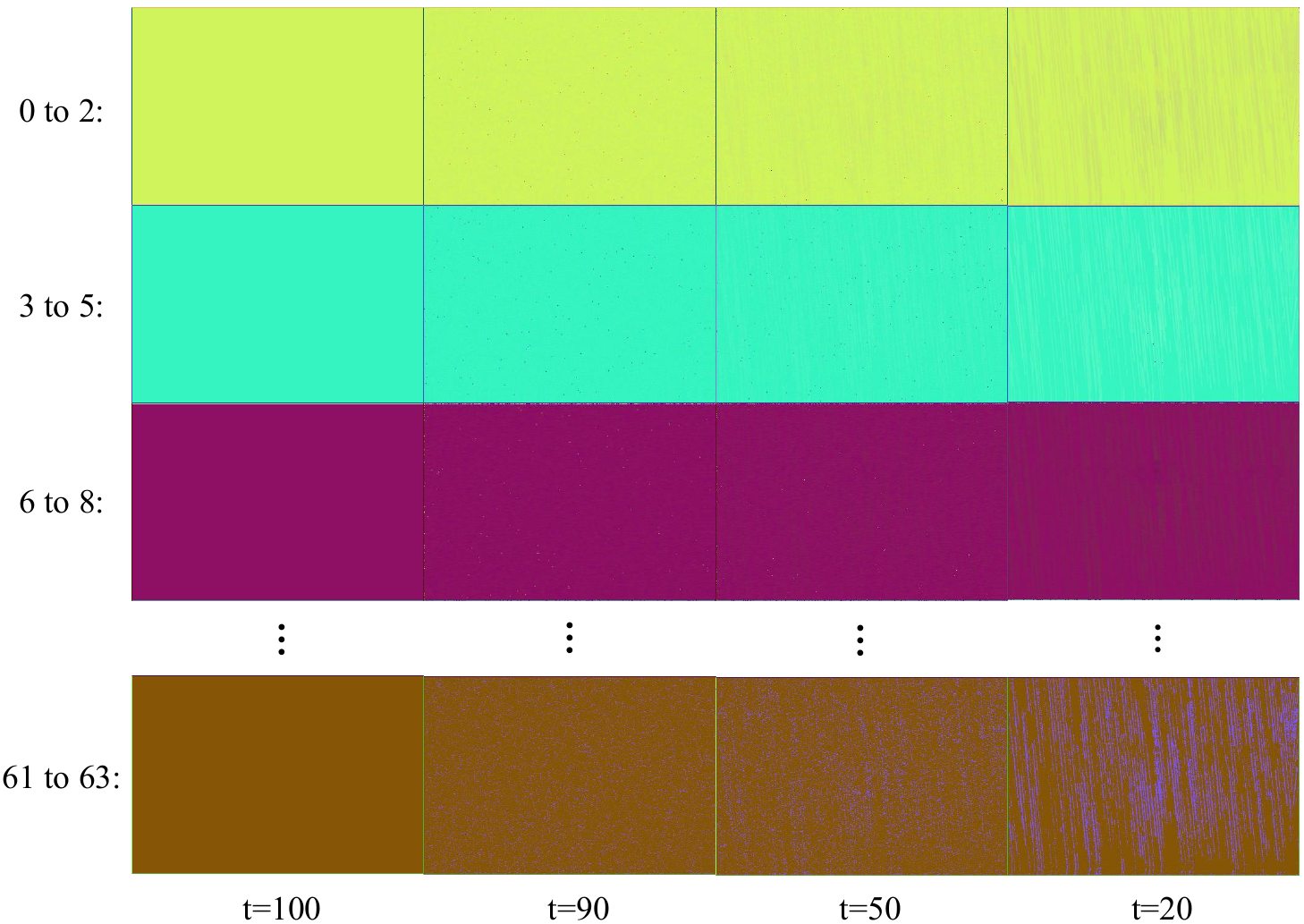}
        \caption{Visualization of DFM features over partial channels and at different times}
        \label{DFM}
    \end{subfigure}
    \vskip\baselineskip
    \begin{subfigure}[b]{0.21\textwidth}
        \centering
        \includegraphics[width=\textwidth]{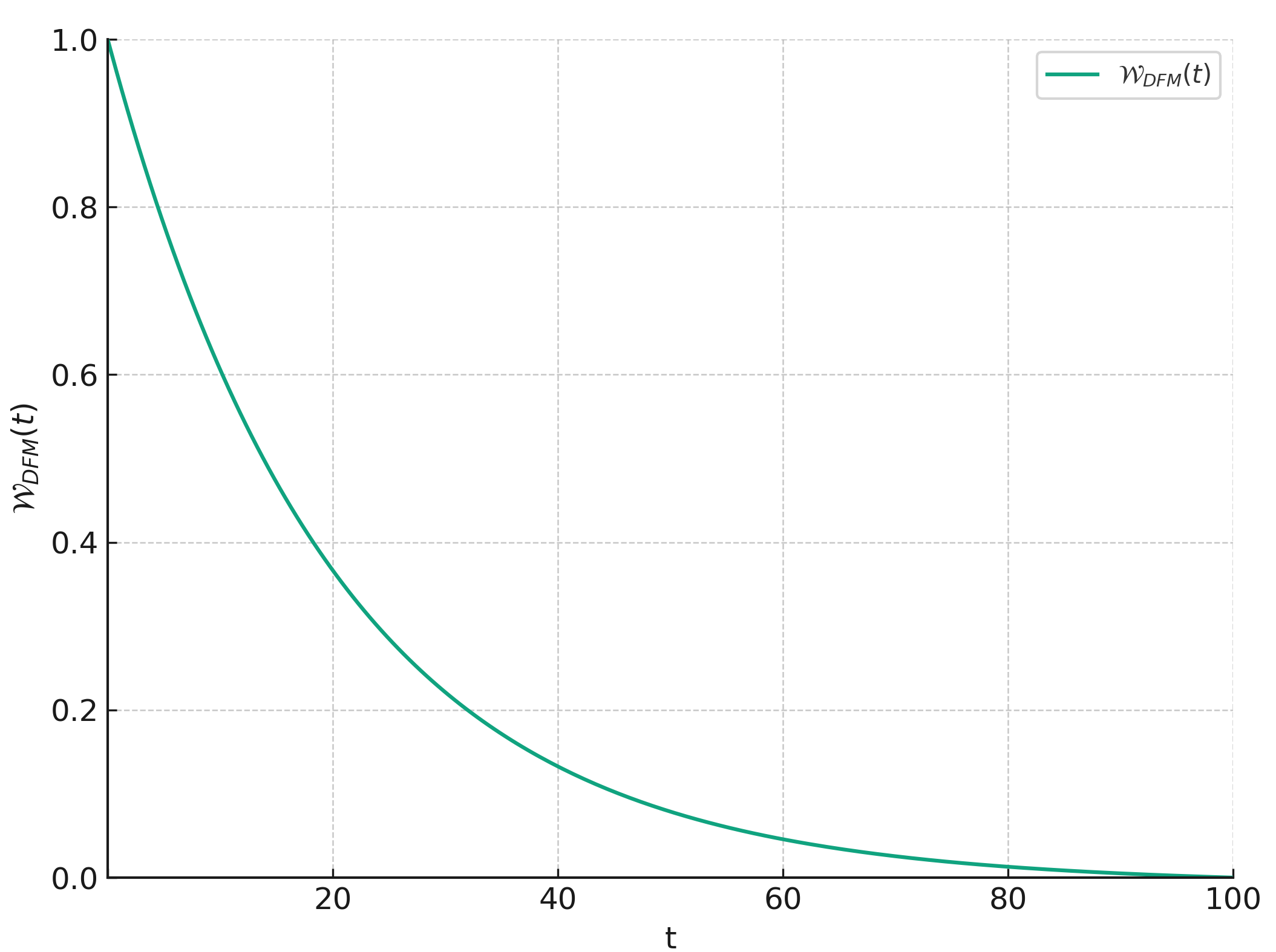}
        \caption{Graph of the weight function on the $t\in[0,100]$ when $a=5$}
        \label{weight_dfm}
    \end{subfigure}
    \hfill
    \begin{subfigure}[b]{0.23\textwidth}
        \centering
        \includegraphics[width=\textwidth]{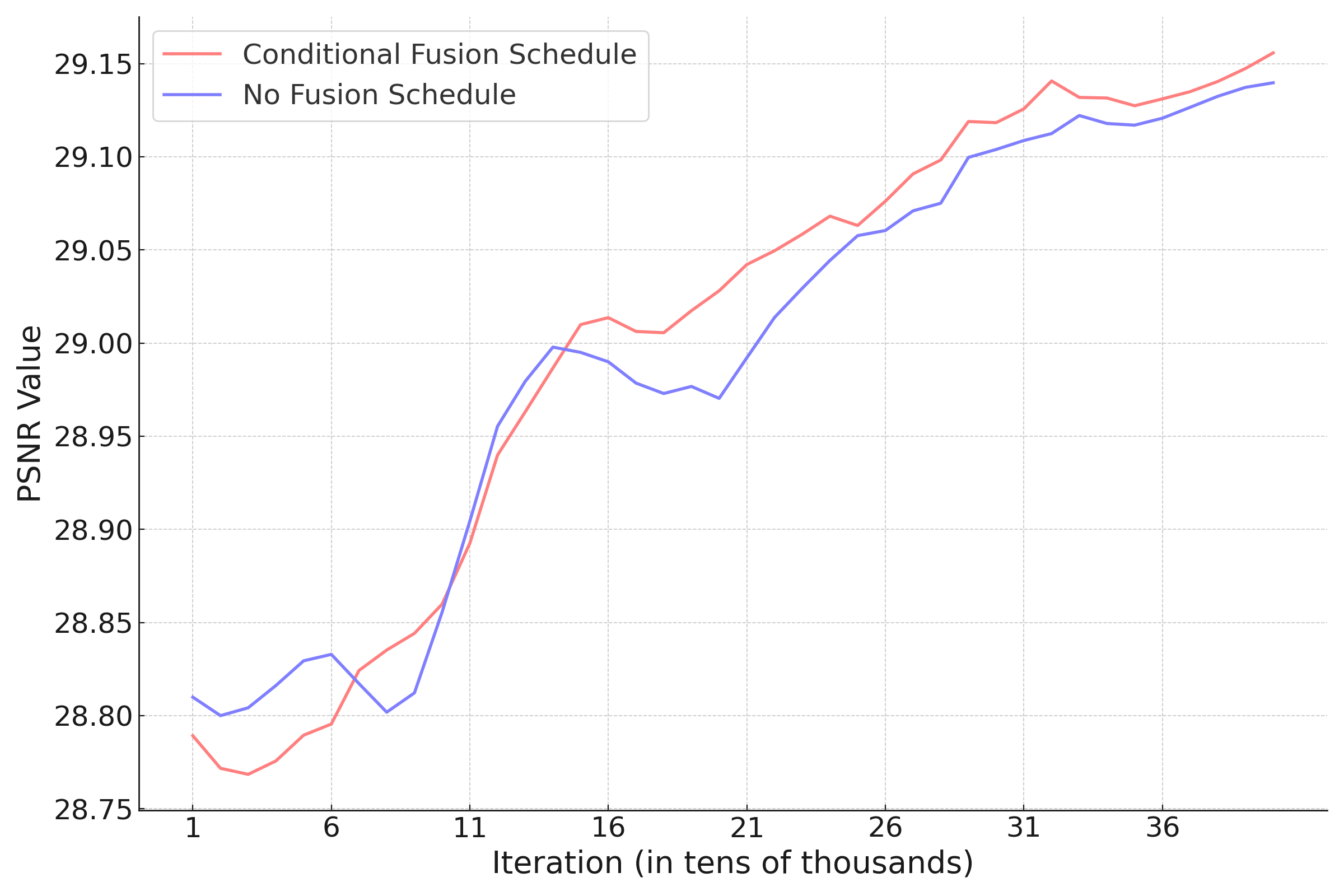}
        \caption{Graph of Conditional Fusion Schedule ablation study on the validation dataset.}
        \label{schedule_comp}
    \end{subfigure}
    \caption{Graph of Conditional Fusion Schedule ablation study on the validation dataset.}
\end{figure}

%% file: content/RelatedWork.tex
\section{Related Works}
Image restoration models based on CNN \cite{dong2014learning, lim2017enhanced} and Transformer \cite{zamir2022restormer, adrai2024deep} aim to utilize pairs of high-quality and low-quality images to learn their degradation process end-to-end, thereby enabling the recovery of high-quality images from low-quality ones. These methods \cite{yan2023cascaded, li2023lsdir, yang2023visual} commonly employ a U-Net architecture, whose hierarchical multi-scale representation and learning via skip connections between shallow and deep layers have been proven to be highly effective in pixel-level end-to-end training, making them well-suited for image restoration tasks.

Diffusion-baed models \cite{xia2023diffir, fei2023generative, zhu2023denoising, luo2023refusion} have also received widespread attention in recent years. The core idea is to embed low-quality images as information within the diffusion process, guiding the model to generate corresponding high-quality images. 
Some approaches \cite{kawar2021snips, chungdiffusion} regard the degradation process as a linear noise inverse process, employing the degradation matrix and conditions for Classifier-Free \cite{dhariwal2021diffusion} conditional control of diffusion models. Other approaches \cite{li2023bbdm, zhoudenoising, liu2023i2sb, yue2023image} model the degradation process as a specialized diffusion process, facilitating the transition of high-quality images into their low-quality ones. This method employs neural networks to predict the noise, starting from the low-quality image and iteratively applying the reverse SDE to generate the high-quality image.

%% file: content/Preliminary.tex
\section{Preliminaries}
In this section, we will primarily introduce the theoretical foundation of GOUB. Let the initial state $\mathbf{x}_0$ represents the HQ image and the final state $\mathbf{x}_T$ represents its corresponding LQ image, the forward process of GOUB can be described as:
{\small
\begin{equation}\label{forward}
\begin{aligned}
\mathrm{d}\mathbf{x}_t=\left(\theta _t + g^2_t \frac{e^{-2\bar{\theta}_{t:T}}}{\bar\sigma_{t:T}^2}  \right)(\mathbf{x}_T - \mathbf{x}_t) \mathrm{d}t + g_t \mathrm{d}\mathbf{w}_t,\\
\bar{\theta}_{t:T} = \int_t^T{\theta _zdz},
\quad \bar \sigma_{t:T}^2=\frac{g^2_T}{2\theta_T}\left( 1-e^{-2\bar{\theta}_{t:T}}\right),
\end{aligned}
\end{equation}
}

where $\theta_t$ is a scalar drift coefficient, $g_t$ denotes the diffusion coefficient and $\mathbf{w}_t$ represents the standard Brownian motion. In addition, we also require that $\theta_t$, $g_t$ satisfy the relationship: $2\lambda^2=g^2_t/\theta_t$, where $\lambda^2$ is a given constant scalar. 
The SDE will definitely pass through the given point $\mathbf{x}_T$ at $t=T$, meaning the marginal distribution $p(\mathbf{x}_T\mid \mathbf{x}_0)=\delta(x_T)$ at that time. This is akin to a bridge connecting the points $\mathbf{x}_0$ and $\mathbf{x}_T$, hence we refer to this type of model as a diffusion bridge model. Correspondingly, the forward process at any given moment $t$ can be defined as follows: 
{\small
\begin{equation}\label{forward_transition}
\begin{gathered}
p(\mathbf{x}_t\mid \mathbf{x}_0, \mathbf{x}_T)
=N(\mathbf{\bar m'}_t, \bar\sigma'^{2}_{t}\mathbf{I}),\\
\mathbf{\bar m'}_t = e^{-\bar{\theta}_{t}}\frac{\bar\sigma_{t:T}^2}{\bar\sigma_{T}^2}\mathbf{x}_0+\left[ \left(1-e^{-\bar{\theta}_{t}}\right)\frac{\bar\sigma_{t:T}^2}{\bar\sigma_{T}^2} + e^{-2\bar{\theta}_{t:T}}\frac{\bar\sigma_{t}^2}{\bar\sigma_{T}^2} \right]\mathbf{x}_T\\
\bar\sigma'^{2}_{t} = \frac{\bar\sigma_{t}^2\bar\sigma_{t:T}^2}{\bar\sigma_{T}^2}
\end{gathered}
\end{equation}
}
The reverse SDE of Equation \eqref{forward} is:
{\small
\begin{equation}\label{reverse}
\begin{aligned}
\mathrm{d}\mathbf{x}_t =& \Bigg[ \left(\theta _t + g^2_t \frac{e^{-2\bar{\theta}_{t:T}}}{\bar\sigma_{t:T}^2}  \right)(\mathbf{x}_T - \mathbf{x}_t) \Bigg. \\
& \Bigg. - g^2_t\nabla_{\mathbf{x}_t}\log p(\mathbf{x}_t\mid \mathbf{x}_T) \Bigg] \mathrm{d}t + g_t \mathrm{d}\mathbf{w}_t,
\end{aligned}
\end{equation}
}
In practice, we set $T=1$ and use 100 interval time steps and Eluer Sampler for sampling. Similar to diffusion models, we can parameterize noise as $\bm\epsilon_{\bm\theta}(\mathbf x_t,\mathbf x_T,t)$ and final training object is:
{\footnotesize
\begin{equation}
\begin{aligned}
\mathcal{L} = &\mathbb{E}_{t,\mathbf x_0,\mathbf x_t,\mathbf x_T}
\left[
\frac{1}{2g_t^2} 
\Bigg\| 
\frac{1}{\bar\sigma'^{2}_{t}}\left[\bar\sigma'^{2}_{t-1}(\mathbf x_t- b \mathbf x_T)a +(\bar\sigma'^{2}_{t}-\bar\sigma'^{2}_{t-1}a^2)\mathbf{\bar m'}_t \right]
\Bigg.\right.\\
&\Bigg.\left.  - \mathbf x_{t} + \left(\theta _t + g^2_t \frac{e^{-2\bar{\theta}_{t:T}}}{\bar\sigma_{t:T}^2}  \right) (\mathbf x_T - \mathbf x_t) + \frac{g^2_t}{\bar\sigma'_{t}}\bm\epsilon_{\bm\theta}(\mathbf x_t,\mathbf x_T,t)
\Bigg\|
\right]
\end{aligned}
\end{equation}
}
The conditional score $\nabla_{\mathbf{x}_t}\log p(\mathbf{x}_t\mid \mathbf{x}_T) \approx \nabla_{\mathbf x_t}\log p_\theta(\mathbf x_t\mid \mathbf x_T) = -\bm\epsilon_{\bm\theta}(\mathbf x_t, \mathbf x_T,t)/\bar\sigma'_{t}$. Therefore, starting from low-quality image $\mathbf x_T$, we can recover $\mathbf x_0$ by utilizing Equation \eqref{reverse} to perform reverse iteration.

\begin{figure}[t]
  \centering
  \includegraphics[width=0.5\textwidth]{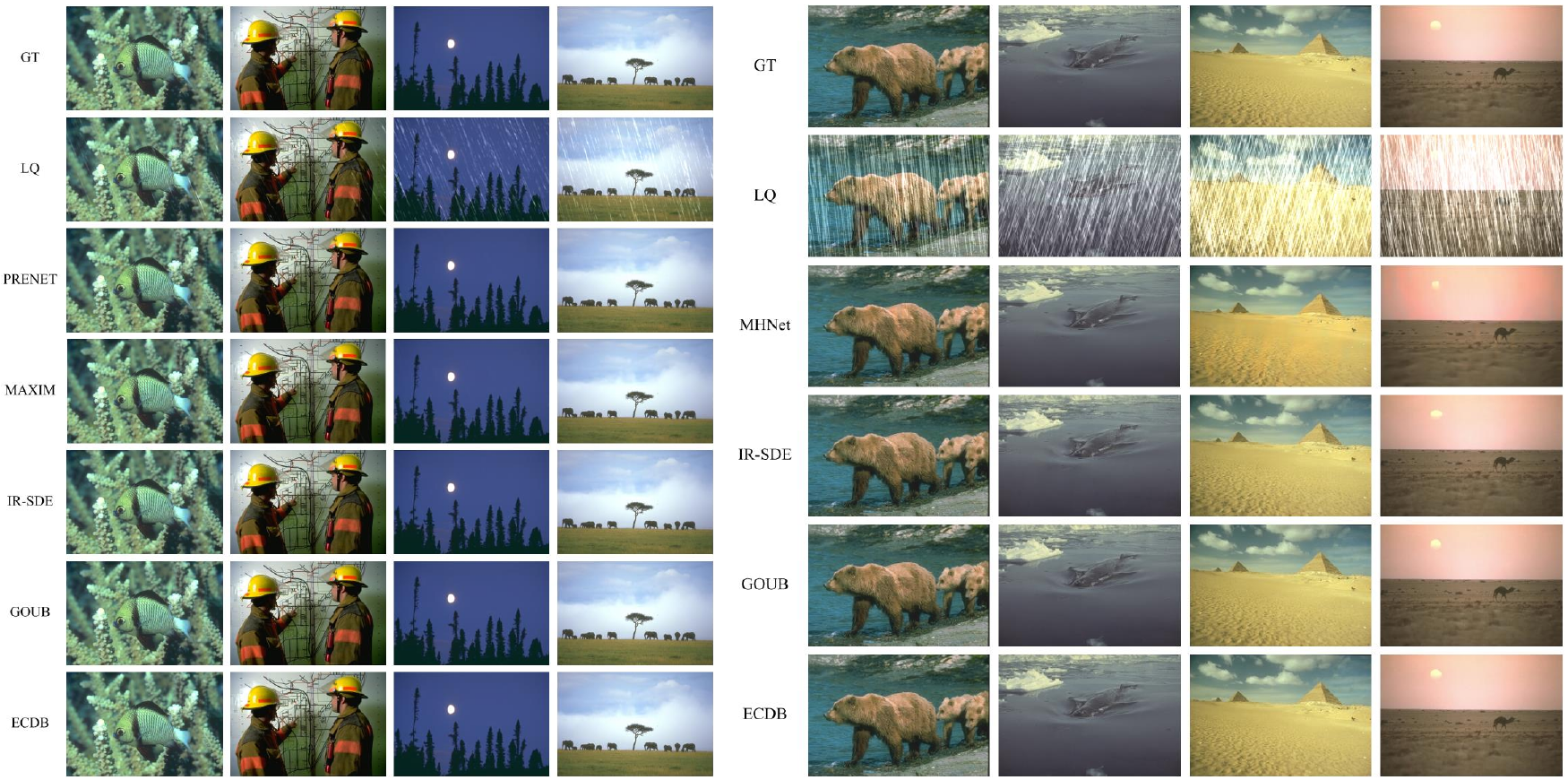}
  \caption{Qualitative comparison of the visual results of different deraining methods on the Rain100L (Left) and Rain100H (Right) dataset.}
  \label{rain}
\end{figure}

\begin{figure}[t]
  \centering
  \includegraphics[width=0.5\textwidth]{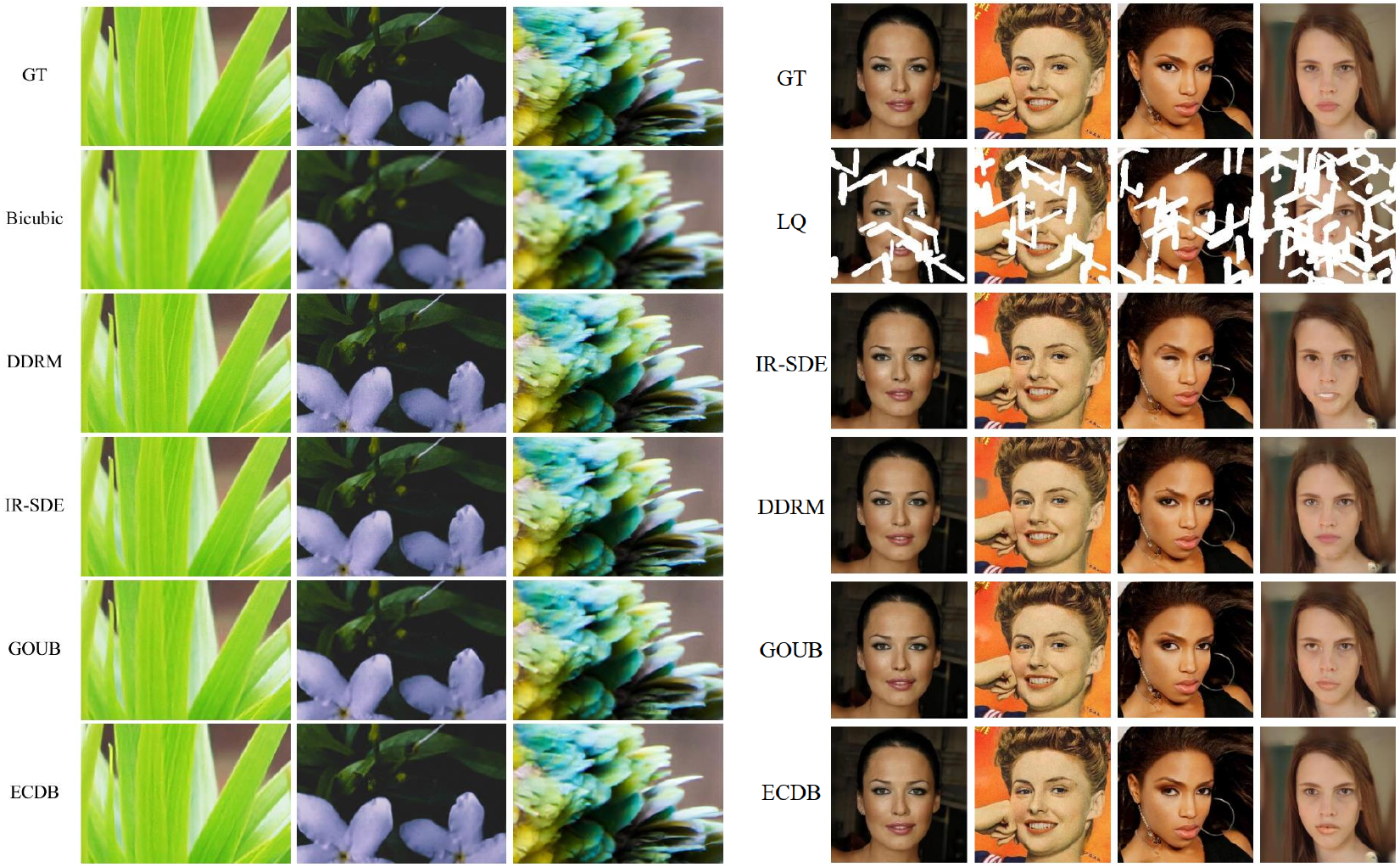}
  \caption{Qualitative comparison of the visual results from different super-resolution methods on the DIV2K dataset (Left) and inpainting methods on the CelebA-HQ dataset (Right).}
  \label{su_in}
\end{figure}

%% file: content/ECDB.tex
\section{ECDB}
\subsection{Architectures}
The most important aspect of the application of diffusion models in the field of image restoration is how to use low-quality images as conditions for precise control of generation, in order to reconstruct the original high-quality images.
Inspired by the ControlNet, we designed and introduced some new conditional control modules to direct the model's generation without altering the original model, making full use of the information from the pretrained model.
As shown in Figure \ref{ECDB}, our ECDB model primarily consists of DM, CHM, DFM and CM, which accept the current state $\mathbf{x}_t$, LQ conditions $\mathbf{x}_T$, and time $t$ as inputs, and output the predicted noise $\epsilon_{\bm \theta}(\mathbf{x}_t, \mathbf{x}_T, t)$ for the $t-1$ step.

DM is the structure used by GOUB, primarily featuring a U-Net architecture that accepts $\mathbf{x}_t$, $\mathbf{x}_T$ and $t$ as inputs. $t$ is first encoded by an MLP encoder, then concatenated with $\mathbf{x}_t$ and $\mathbf{x}_T$ put into the U-Net for processing. Notably, during the upsampling process, it also needs to receive inputs from the corresponding blocks of CM, ensuring the full integration and exchange of conditional information. Additionally, it is crucial that we initialize DM using a pretrained model and freeze parameters during the training process. This approach ensures that DM retains its fundamental denoising ability while solely learning the conditional control function.

The purpose of CHM is to extract LQ condition features. As shown in Figure \ref{CHM}, it primarily focuses on color and edge features across various channels. Its core architecture is composed of four convolution layers and one zero convolution layer that is initialized to zero, with SiLU (Sigmoid Linear Unit) activation function.

DFM is designed to extract degradation features, which correspond to raindrop information in the task of deraining or mask information in the task of inpainting. Its core structure is consistent with the CHM. As shown in Figure \ref{DFM}, we visualized the output features of DFM on different channels for the task of deraining. It is evident that as $t$ approaches 0, the extracted features distinctly resemble raindrops, which further facilitates enhanced conditional control for handling specific tasks.

CM is used to process and integrate conditional feature information. Its upper half structure is consistent with the upsampling module of DM's U-Net and is initialized by it, while the lower half is composed of zero convolution layers.

The aforementioned structures and initialization method ensure that the initial impact of the conditional control part on the denoising process is zero, starting from zero to approach the optimal parameters making the training more stable.

\subsection{Conditional Fusion Schedule}
Reviewing Figure \ref{DFM}, we visualized the degradation features of DFM across different channels at various moments in time. It is evident that at the beginning of the restoration process, when $t$ from 100 to 50, the output consists mostly of meaningless zero information and slight noise, while information about the degradation operator only appears after $t=50$. This is because $\mathbf{x}_t$ has a low level of denoising at the beginning, making $\mathbf{x}_T - \mathbf{x}_t$ close to zero, lacking effective information, and thus it is difficult to perform any control function. However, this part of the ineffective information will also be involved in the training process. Therefore, we propose a feature fusion strategy intended to assign appropriate weights to the DFM features, thereby incorporating them into the fusion process. Our requirement for the weight function $\mathcal{W}_{DFM}$ is that it smoothly approaches 0 from $t=100$ to $50$, and tends towards 1 from $t=50$ to $0$. Considering the natural tendency advantage of exponential functions, we could set it as follows:
\begin{equation}\label{weight}
\mathcal{W}_{DFM}(t) = e^{-at}-e^{-a}t
\end{equation}
In all experiments, we set the parameter $a = 5$. As shown in Figure \ref{weight_dfm}, the weight function is monotonically decreasing, approaching zero almost entirely after $t=50$, thus reducing the impact of corresponding samples on the parameters of DFM update during training. Therefore, the final fusion schedule is as follows:
\begin{equation}\label{fusionschedule}
\textit{Fusion} = \mathbf{x}_t + CHM(\mathbf{x}_T) + \mathcal{W}_{DFM}(t) * DFM(\mathbf{x}_T-\mathbf{x}_t)
\end{equation}

\begin{table}[!ht]
  \centering
  \caption{\textbf{Image Deraining}  Qualitative comparison with the relevant baselines on Rain100H.}
  \label{tbderainh}
  \begin{tabular}{ccccc}
    \toprule
    \textbf{METHOD} & \textbf{PSNR\(\uparrow\)} & \textbf{SSIM\(\uparrow\)} & \textbf{LPIPS\(\downarrow\)} & \textbf{FID\(\downarrow\)} \\
    \midrule
    MHNet & 31.08 & 0.8990 & 0.126 & 57.93\\
    IR-SDE & 31.65 & 0.9041 & 0.047 & 18.64 \\
    GOUB & 31.96 & 0.9028 & 0.046 & 18.14 \\
    \midrule
    ECDB & \textbf{32.23} & \textbf{0.9058} & \textbf{0.043} & \textbf{16.79} \\
    \bottomrule
  \end{tabular}
\end{table}

\begin{table}[!ht]
  \centering
  \caption{\textbf{Image Deraining.}  Qualitative comparison with the relevant baselines on Rain100L.}
  \label{tbderainl}
  \begin{tabular}{ccccc}
    \toprule
    \textbf{METHOD} & \textbf{PSNR\(\uparrow\)} & \textbf{SSIM\(\uparrow\)} & \textbf{LPIPS\(\downarrow\)} & \textbf{FID\(\downarrow\)} \\
    \midrule
    PRENET & 37.48 & 0.9792 & 0.0207 & 10.9 \\
    MAXIM & 38.06 & 0.9770 & 0.0483 & 19.0 \\
    IRSDE & 38.30 & 0.9805 & 0.0141 & 7.94 \\
    GOUB & 39.79 & 0.9830 & 0.0096 & 5.18 \\
    \midrule
    ECDB & \textbf{40.05} & \textbf{0.9836} & \textbf{0.0090} & \textbf{4.56} \\
    \bottomrule
  \end{tabular}
\end{table}

Therefore, DFM only undergoes significant updates when the degradation features are more pronounced, without being affected by ineffective information, resulting in better training outcomes.

%% file: content/Experiment.tex
\section{Experiments}
Our experiments primarily focus on tasks related to deraining, inpainting, and super-resolution, with the evaluation metrics being Peak Signal-to-Noise Ratio (PSNR), Structural Similarity Index (SSIM) \cite{wang2004image}, Learned Perceptual Image Patch Similarity (LPIPS) \cite{zhang2018unreasonable}, and Fréchet Inception Distance (FID) \cite{heusel2017gans}, similar to those used for general image restoration models. Additionally, we conduct ablation studies on various modules and the proposed conditional fusion schedule to demonstrate the advancements of our proposed improvements. Hyperparameters such as \( f_t \), \( \lambda^2 \), and others are set consistently with GOUB. The total training steps are 400 thousand with the initial learning rate set to $2\times 10^{-5}$ and is halved at iterations 100, 200 and 300 thousand.

\subsection{Specific Tasks}
\paragraph{Image Deraining} For the deraining task, we conduct experiments on the Rain100H and Rain100L datasets \cite{yang2017deep} with both 1800 training set and 100 test set. It's important to note that, similar to other deraining models, PSNR and SSIM are computed on the Y channel (YCbCr space). We report state-of-the-art baselines for comparison: MHNet \cite{gao2023mixed}, IR-SDE \cite{luo2023image} and GOUB \cite{yue2023image}. The experimental results are presented in Figure \ref{rain} and Table \ref{tbderainh}, \ref{tbderainl}. From the table, it can be seen that our model surpasses the baselines in all metrics, achieving state-of-the-art (SOTA) results; from the images, it is evident that our model also performs better in terms of details.

\paragraph{Image Super-Resolution} We conducted training and evaluation on the DIV2K validation set for 4$\times$ upscaling \cite{agustsson2017ntire} and all low-resolution images were bicubically rescaled to the same size as their corresponding high-resolution images. To show that our models are in line with the state-of-the-art, we compare to the DDRM \cite{kawar2022denoising}, IR-SDE \cite{luo2023image} and GOUB \cite{yue2023image}. The relevant experimental results are provided in Figure \ref{su_in} and Table \ref{tbsr}. In the 4x super-resolution task, our model achieved SOTA results on the diffusion benchmark models, significantly surpassing previous metrics and also showing superior performance in generation.

\paragraph{Image Inpainting} We have selected the CelebA-HQ $256 \times 256$ datasets \cite{karras2017progressive} for both training and testing with 100 thin masks. We compare our models with several current baseline inpainting approaches such as DDRM \cite{kawar2022denoising}, IR-SDE \cite{luo2023image} and GOUB \cite{yue2023image}. The relevant experimental results are shown in Figure \ref{su_in} and Table \ref{tbinpaint}. Similarly, for the inpainting task, our model also achieved SOTA results, performing exceptionally well in details such as the background and eyes.

\begin{table}[!t]
  \centering
  \caption{\textbf{Image Inpainting.} Qualitative comparison with the relevant baselines on CelebA-HQ.}
  \label{tbinpaint}
  \begin{tabular}{ccccc}
    \toprule
    \textbf{METHOD} & \textbf{PSNR\(\uparrow\)} & \textbf{SSIM\(\uparrow\)} & \textbf{LPIPS\(\downarrow\)} & \textbf{FID\(\downarrow\)} \\
    \midrule
    DDRM & 27.16 & 0.8993 & 0.089 & 37.02\\
    IR-SDE & 28.37 & 0.9166 & 0.046 & 25.13 \\
    GOUB & 28.98 & 0.9067 & 0.0378 & 4.30 \\
    \midrule
    ECDB & \textbf{29.01} & \textbf{0.9071} & \textbf{0.0372} & \textbf{4.13} \\
    \bottomrule
  \end{tabular}
\end{table}

\begin{table}[!t]
  \centering
  \caption{\textbf{Image 4$\times$ Super-Resolution.} Qualitative comparison with the relevant baselines on DIV2K.}
  \label{tbsr}
  \begin{tabular}{ccccc}
    \toprule
    \textbf{METHOD} & \textbf{PSNR\(\uparrow\)} & \textbf{SSIM\(\uparrow\)} & \textbf{LPIPS\(\downarrow\)} & \textbf{FID\(\downarrow\)} \\
    \midrule
    DDRM & 24.35 & 0.5927 & 0.364 & 78.71\\
    IR-SDE & 25.90 & 0.6570 & 0.231 & 45.36 \\
    GOUB & 26.89 & 0.7478 & 0.220 & 20.85 \\
    \midrule
    ECDB & \textbf{27.39} & \textbf{0.7682} & \textbf{0.212} & \textbf{18.88} \\
    \bottomrule
  \end{tabular}
\end{table}

\subsection{Ablation Study}
To explore the effects of CHM, DFM, and conditional fusion schedule, we compared the impact of different choices on the final results for the deraining task (Rain100H). For the experiment without using the Conditional Fusion Schedule, we directly add all the conditional features.
The experimental results are shown in the Table \ref{ablation}.

\begin{table}[!ht]
\caption{Ablation study on deraining task with Rain100H dataset.}
\label{ablation}
\resizebox{\columnwidth}{!}{
\begin{tabular}{ccc|cccc}
\toprule
\textbf{CHM} & \textbf{DFM} & \textbf{Fusion Schedule} & \textbf{PSNR\(\uparrow\)} & \textbf{SSIM\(\uparrow\)} & \textbf{LPIPS\(\downarrow\)} & \textbf{FID\(\downarrow\)} \\
\midrule
 $\checkmark$   &      &         & 
32.07 & 0.9036 & 0.046 & 18.07 \\
    &   $\checkmark$   &         & 
31.90 & 0.9025 & 0.048 & 19.22\\
 $\checkmark$   &   $\checkmark$   &        & 
32.18 & 0.9056 & 0.044 & 16.98 \\
  $\checkmark$   &   $\checkmark$   &    $\checkmark$     & 
\textbf{32.23} & \textbf{0.9058} & \textbf{0.043} & \textbf{16.79} \\
\bottomrule
\end{tabular}
}
\end{table}



%% file: content/Conclusion.tex
\section{Conclusion}
In this paper, we focus on the task of image restoration and aim to conditionally enhance control of the original diffusion bridge model by proposing the ECDB model, which consists of DM, CHM, DFM, and CM. DM is the original denoising model, maintaining unchanged parameters, primarily responsible for predicting noise. CHM, DFM, and CM are enhanced conditional control modules used to handle low-quality image information as conditions. Additionally, we address the characteristic of low denoising at larger values of $t$ in the diffusion process, proposing a Conditional Fusion Schedule for conditional fusion of features extracted from each module. Ultimately, our model achieves state-of-the-art results on many image restoration tasks, such as deraining, inpainting and super-resolution.

%% file: main.bbl
\begin{thebibliography}{10}
\providecommand{\url}[1]{#1}
\csname url@samestyle\endcsname
\providecommand{\newblock}{\relax}
\providecommand{\bibinfo}[2]{#2}
\providecommand{\BIBentrySTDinterwordspacing}{\spaceskip=0pt\relax}
\providecommand{\BIBentryALTinterwordstretchfactor}{4}
\providecommand{\BIBentryALTinterwordspacing}{\spaceskip=\fontdimen2\font plus
\BIBentryALTinterwordstretchfactor\fontdimen3\font minus \fontdimen4\font\relax}
\providecommand{\BIBforeignlanguage}[2]{{%
\expandafter\ifx\csname l@#1\endcsname\relax
\typeout{** WARNING: IEEEtran.bst: No hyphenation pattern has been}%
\typeout{** loaded for the language `#1'. Using the pattern for}%
\typeout{** the default language instead.}%
\else
\language=\csname l@#1\endcsname
\fi
#2}}
\providecommand{\BIBdecl}{\relax}
\BIBdecl

\bibitem{banham1997digital}
M.~R. Banham and A.~K. Katsaggelos, ``Digital image restoration,'' \emph{IEEE signal processing magazine}, vol.~14, no.~2, pp. 24--41, 1997.

\bibitem{zhou1988image}
Y.-T. Zhou, R.~Chellappa, A.~Vaid, and B.~K. Jenkins, ``Image restoration using a neural network,'' \emph{IEEE transactions on acoustics, speech, and signal processing}, vol.~36, no.~7, pp. 1141--1151, 1988.

\bibitem{tian2023multi}
C.~Tian, M.~Zheng, W.~Zuo, B.~Zhang, Y.~Zhang, and D.~Zhang, ``Multi-stage image denoising with the wavelet transform,'' \emph{Pattern Recognition}, vol. 134, p. 109050, 2023.

\bibitem{chen2023ffti}
Y.~Chen, R.~Xia, K.~Zou, and K.~Yang, ``Ffti: Image inpainting algorithm via features fusion and two-steps inpainting,'' \emph{Journal of Visual Communication and Image Representation}, vol.~91, p. 103776, 2023.

\bibitem{corneanu2024latentpaint}
C.~Corneanu, R.~Gadde, and A.~M. Martinez, ``Latentpaint: Image inpainting in latent space with diffusion models,'' in \emph{Proceedings of the IEEE/CVF Winter Conference on Applications of Computer Vision}, 2024, pp. 4334--4343.

\bibitem{tao2024mffdnet}
W.~Tao, X.~Yan, Y.~Wang, and M.~Wei, ``Mffdnet: Single image deraining via dual-channel mixed feature fusion,'' \emph{IEEE Transactions on Instrumentation and Measurement}, 2024.

\bibitem{chen2023dual}
Z.~Chen, Y.~Zhang, J.~Gu, L.~Kong, X.~Yang, and F.~Yu, ``Dual aggregation transformer for image super-resolution,'' in \emph{Proceedings of the IEEE/CVF international conference on computer vision}, 2023, pp. 12\,312--12\,321.

\bibitem{yue2024resshift}
Z.~Yue, J.~Wang, and C.~C. Loy, ``Resshift: Efficient diffusion model for image super-resolution by residual shifting,'' \emph{Advances in Neural Information Processing Systems}, vol.~36, 2024.

\bibitem{sohl2015deep}
J.~Sohl-Dickstein, E.~Weiss, N.~Maheswaranathan, and S.~Ganguli, ``Deep unsupervised learning using nonequilibrium thermodynamics,'' in \emph{International conference on machine learning}.\hskip 1em plus 0.5em minus 0.4em\relax PMLR, 2015, pp. 2256--2265.

\bibitem{ho2020denoising}
J.~Ho, A.~Jain, and P.~Abbeel, ``Denoising diffusion probabilistic models,'' \emph{Advances in neural information processing systems}, vol.~33, pp. 6840--6851, 2020.

\bibitem{songscore}
Y.~Song, J.~Sohl-Dickstein, D.~P. Kingma, A.~Kumar, S.~Ermon, and B.~Poole, ``Score-based generative modeling through stochastic differential equations,'' in \emph{International Conference on Learning Representations}, 2021.

\bibitem{kawar2021snips}
B.~Kawar, G.~Vaksman, and M.~Elad, ``Snips: Solving noisy inverse problems stochastically,'' \emph{Advances in Neural Information Processing Systems}, vol.~34, pp. 21\,757--21\,769, 2021.

\bibitem{chungdiffusion}
H.~Chung, J.~Kim, M.~T. Mccann, M.~L. Klasky, and J.~C. Ye, ``Diffusion posterior sampling for general noisy inverse problems,'' in \emph{The Eleventh International Conference on Learning Representations}, 2023.

\bibitem{chung2022improving}
H.~Chung, B.~Sim, D.~Ryu, and J.~C. Ye, ``Improving diffusion models for inverse problems using manifold constraints,'' \emph{Advances in Neural Information Processing Systems}, vol.~35, pp. 25\,683--25\,696, 2022.

\bibitem{li2023bbdm}
B.~Li, K.~Xue, B.~Liu, and Y.-K. Lai, ``Bbdm: Image-to-image translation with brownian bridge diffusion models,'' in \emph{Proceedings of the IEEE/CVF Conference on Computer Vision and Pattern Recognition}, 2023, pp. 1952--1961.

\bibitem{liu2023i2sb}
G.-H. Liu, A.~Vahdat, D.-A. Huang, E.~A. Theodorou, W.~Nie, and A.~Anandkumar, ``I2sb: image-to-image schr{\"o}dinger bridge,'' in \emph{Proceedings of the 40th International Conference on Machine Learning}, 2023, pp. 22\,042--22\,062.

\bibitem{shi2024diffusion}
Y.~Shi, V.~De~Bortoli, A.~Campbell, and A.~Doucet, ``Diffusion schr{\"o}dinger bridge matching,'' \emph{Advances in Neural Information Processing Systems}, vol.~36, 2024.

\bibitem{tongsimulation}
A.~Tong, N.~Malkin, K.~FATRAS, L.~Atanackovic, Y.~Zhang, G.~Huguet, G.~Wolf, and Y.~Bengio, ``Simulation-free schr{\"o}dinger bridges via score and flow matching,'' in \emph{ICML Workshop on New Frontiers in Learning, Control, and Dynamical Systems}, 2023.

\bibitem{zhoudenoising}
L.~Zhou, A.~Lou, S.~Khanna, and S.~Ermon, ``Denoising diffusion bridge models,'' in \emph{The Twelfth International Conference on Learning Representations}, 2024.

\bibitem{yue2023image}
C.~Yue, Z.~Peng, J.~Ma, S.~Du, P.~Wei, and D.~Zhang, ``Image restoration through generalized ornstein-uhlenbeck bridge,'' in \emph{Proceedings of the 41st International Conference on Machine Learning}, 2024.

\bibitem{dong2014learning}
C.~Dong, C.~C. Loy, K.~He, and X.~Tang, ``Learning a deep convolutional network for image super-resolution,'' in \emph{Computer Vision--ECCV 2014: 13th European Conference, Zurich, Switzerland, September 6-12, 2014, Proceedings, Part IV 13}.\hskip 1em plus 0.5em minus 0.4em\relax Springer, 2014, pp. 184--199.

\bibitem{lim2017enhanced}
B.~Lim, S.~Son, H.~Kim, S.~Nah, and K.~Mu~Lee, ``Enhanced deep residual networks for single image super-resolution,'' in \emph{Proceedings of the IEEE conference on computer vision and pattern recognition workshops}, 2017, pp. 136--144.

\bibitem{zamir2022restormer}
S.~W. Zamir, A.~Arora, S.~Khan, M.~Hayat, F.~S. Khan, and M.-H. Yang, ``Restormer: Efficient transformer for high-resolution image restoration,'' in \emph{Proceedings of the IEEE/CVF conference on computer vision and pattern recognition}, 2022, pp. 5728--5739.

\bibitem{adrai2024deep}
T.~Adrai, G.~Ohayon, M.~Elad, and T.~Michaeli, ``Deep optimal transport: A practical algorithm for photo-realistic image restoration,'' \emph{Advances in Neural Information Processing Systems}, vol.~36, 2024.

\bibitem{yan2023cascaded}
L.~Yan, M.~Zhao, S.~Liu, S.~Shi, and J.~Chen, ``Cascaded transformer u-net for image restoration,'' \emph{Signal Processing}, vol. 206, p. 108902, 2023.

\bibitem{li2023lsdir}
Y.~Li, K.~Zhang, J.~Liang, J.~Cao, C.~Liu, R.~Gong, Y.~Zhang, H.~Tang, Y.~Liu, D.~Demandolx \emph{et~al.}, ``Lsdir: A large scale dataset for image restoration,'' in \emph{Proceedings of the IEEE/CVF Conference on Computer Vision and Pattern Recognition}, 2023, pp. 1775--1787.

\bibitem{yang2023visual}
Z.~Yang, J.~Huang, J.~Chang, M.~Zhou, H.~Yu, J.~Zhang, and F.~Zhao, ``Visual recognition-driven image restoration for multiple degradation with intrinsic semantics recovery,'' in \emph{Proceedings of the IEEE/CVF Conference on Computer Vision and Pattern Recognition}, 2023, pp. 14\,059--14\,070.

\bibitem{xia2023diffir}
B.~Xia, Y.~Zhang, S.~Wang, Y.~Wang, X.~Wu, Y.~Tian, W.~Yang, and L.~Van~Gool, ``Diffir: Efficient diffusion model for image restoration,'' in \emph{Proceedings of the IEEE/CVF International Conference on Computer Vision}, 2023, pp. 13\,095--13\,105.

\bibitem{fei2023generative}
B.~Fei, Z.~Lyu, L.~Pan, J.~Zhang, W.~Yang, T.~Luo, B.~Zhang, and B.~Dai, ``Generative diffusion prior for unified image restoration and enhancement,'' in \emph{Proceedings of the IEEE/CVF Conference on Computer Vision and Pattern Recognition}, 2023, pp. 9935--9946.

\bibitem{zhu2023denoising}
Y.~Zhu, K.~Zhang, J.~Liang, J.~Cao, B.~Wen, R.~Timofte, and L.~Van~Gool, ``Denoising diffusion models for plug-and-play image restoration,'' in \emph{Proceedings of the IEEE/CVF Conference on Computer Vision and Pattern Recognition}, 2023, pp. 1219--1229.

\bibitem{luo2023refusion}
Z.~Luo, F.~K. Gustafsson, Z.~Zhao, J.~Sj{\"o}lund, and T.~B. Sch{\"o}n, ``Refusion: Enabling large-size realistic image restoration with latent-space diffusion models,'' in \emph{Proceedings of the IEEE/CVF conference on computer vision and pattern recognition}, 2023, pp. 1680--1691.

\bibitem{dhariwal2021diffusion}
P.~Dhariwal and A.~Nichol, ``Diffusion models beat gans on image synthesis,'' \emph{Advances in neural information processing systems}, vol.~34, pp. 8780--8794, 2021.

\bibitem{wang2004image}
Z.~Wang, A.~C. Bovik, H.~R. Sheikh, and E.~P. Simoncelli, ``Image quality assessment: from error visibility to structural similarity,'' \emph{IEEE transactions on image processing}, vol.~13, no.~4, pp. 600--612, 2004.

\bibitem{zhang2018unreasonable}
R.~Zhang, P.~Isola, A.~A. Efros, E.~Shechtman, and O.~Wang, ``The unreasonable effectiveness of deep features as a perceptual metric,'' in \emph{Proceedings of the IEEE conference on computer vision and pattern recognition}, 2018, pp. 586--595.

\bibitem{heusel2017gans}
M.~Heusel, H.~Ramsauer, T.~Unterthiner, B.~Nessler, and S.~Hochreiter, ``Gans trained by a two time-scale update rule converge to a local nash equilibrium,'' \emph{Advances in neural information processing systems}, vol.~30, 2017.

\bibitem{yang2017deep}
W.~Yang, R.~T. Tan, J.~Feng, J.~Liu, Z.~Guo, and S.~Yan, ``Deep joint rain detection and removal from a single image,'' in \emph{Proceedings of the IEEE conference on computer vision and pattern recognition}, 2017, pp. 1357--1366.

\bibitem{gao2023mixed}
H.~Gao and D.~Dang, ``Mixed hierarchy network for image restoration,'' \emph{arXiv preprint arXiv:2302.09554}, 2023.

\bibitem{luo2023image}
Z.~Luo, F.~K. Gustafsson, Z.~Zhao, J.~Sj{\"o}lund, and T.~B. Sch{\"o}n, ``Image restoration with mean-reverting stochastic differential equations,'' in \emph{Proceedings of the 40th International Conference on Machine Learning}, 2023, pp. 23\,045--23\,066.

\bibitem{agustsson2017ntire}
E.~Agustsson and R.~Timofte, ``Ntire 2017 challenge on single image super-resolution: Dataset and study,'' in \emph{Proceedings of the IEEE conference on computer vision and pattern recognition workshops}, 2017, pp. 126--135.

\bibitem{kawar2022denoising}
B.~Kawar, M.~Elad, S.~Ermon, and J.~Song, ``Denoising diffusion restoration models,'' \emph{Advances in Neural Information Processing Systems}, vol.~35, pp. 23\,593--23\,606, 2022.

\bibitem{karras2017progressive}
T.~Karras, T.~Aila, S.~Laine, and J.~Lehtinen, ``Progressive growing of gans for improved quality, stability, and variation,'' \emph{arXiv preprint arXiv:1710.10196}, 2017.

\end{thebibliography}
